\documentclass[twocolumn,dvipdfmx]{aastex631}

\usepackage{ulem}
\usepackage{xcolor}
\DeclareRobustCommand{\ogidel}{\bgroup\markoverwith{\textcolor{magenta}{\rule[.5ex]{2pt}{0.4pt}}}\ULon}

\begin{document}

\title{Evolution of a Water-rich Atmosphere Formed by a Giant Impact on an Earth-sized Planet}

\correspondingauthor{Kenji Kurosaki}
\author[0000-0002-0786-7307]{Kenji Kurosaki}
\affiliation{Graduate School of Science, Kobe University, 1-1, Rokkkodai-cho, Nada-ku, Kobe, Hyogo 657-8501, Japan}
\email{kenji.kurosaki@people.kobe-u.ac.jp}

\author[0000-0003-4676-0251]{Yasunori Hori}
\affiliation{National Astronomical Observatory of Japan, 2-21-1, Osawa, Mitaka, 181-8588 Tokyo, Japan}
\affiliation{Astrobiology Center, 2-21-1, Osawa, Mitaka, 181-8588 Tokyo, Japan}

\author[0000-0002-8300-7990]{Masahiro Ogihara}
\affiliation{Tsung-Dao Lee Institute, Shanghai Jiao Tong University, 520 Shengrong Road, Shanghai 201210, China}
\affiliation{Earth-Life Science Institute, Tokyo Institute of Technology, Meguro, Tokyo 152-8550, Japan}

\author[0000-0002-1932-3358]{Masanobu Kunitomo}
\affiliation{Department of Physics, Kurume University, 67 Asahi-machi, Kurume, Fukuoka 830-0011, Japan}



\begin{abstract}
The atmosphere of a terrestrial planet that is replenished with secondary gases should have accumulated hydrogen-rich gas from its protoplanetary disk. Although a giant impact blows off a large fraction of the primordial atmosphere of a terrestrial planet in the late formation stage, the remaining atmosphere can become water-rich via chemical reactions between hydrogen and vaporized core material.
We find that a water-rich post-impact atmosphere forms when a basaltic or CI chondrite core is assumed. In contrast, little post-impact water is generated for an enstatite chondrite core.
We investigate the X-ray- and UV-driven mass loss from an Earth-mass planet with an impact-induced multi-component H$_2$--He--H$_2$O atmosphere for Gyrs. We show that water is left in the atmosphere of an Earth-mass planet when the low flux of escaping hydrogen cannot drag water upward via collisions. For a water-dominated atmosphere to form, the atmospheric mass fraction of an Earth-mass planet with an oxidizing core after a giant impact must be less than a few times $0.1$\%.
We also find that Earth-mass planets with water-dominated atmospheres can exist at semimajor axes ranging from a few times 0.1\,au to a few\,au around a Sun-like star depending on the mass loss efficiency. 
Such planets are important targets for atmospheric characterization in the era of JWST. Our results indicate that efficient mixing between hydrogen and rocky components during giant impacts can play a role in the production of water in an Earth-mass planet.
\end{abstract}

\keywords{Planetary interior (1248), Exoplanet atmospheric evolution (2308)}


\section{Introduction} \label{sec:intro}
Exoplanets as small as Earth are now being discovered by ground-based and spaceborne telescopes.
In the proximity of a central star with orbital periods of less than about 100 days, small planets with radii of $\leq$3\,$R_\oplus$ are more common than Neptune-sized or larger planets.
\citep{2013PNAS..11019273P, 2015ApJ...807...45D}. 
Statistical studies on the population of small planets from {\it Kepler} found that the intriguing transition between bare rocky planets and planets with atmospheres of typically $<$10\,wt\% \citep{2014ApJ...783L...6W, 2015ApJ...801...41R}, 
known as the radius gap, lies around 1.5--2\,$R_\oplus$ \citep{2018AJ....156..264F, 2018AJ....155..127H}. 
Also, the mass-radius relationships of exoplanets with radii smaller than $1.5\,R_\oplus$ suggests that they may have rocky compositions similar to that of Earth \citep[e.g.][]{2019PNAS..116.9723Z}.

This study focuses on Earth-sized planets.
Earth-sized rocky planets possess only thin or nonexistent atmospheres \citep{2015ApJ...801...41R},
although they may have accumulated dense atmospheres in the stage of planet formation.
The primordial atmosphere of a terrestrial planet originates from nebular gas rich in H$_2$/He-rich \citep{Ikoma2012, Lee2018}.
A large fraction of the H$_2$/He atmosphere can be lost by a giant impact event.
A head-on collision causes atmospheric blow-off when the velocity of the ground motion resulting from a strong shock exceeds the escape velocity on the planetary surface \citep{Ahrens1993,Genda2003,Schlichting2015}. 
In contrast, oblique impacts that frequently occur in the giant impact stage remove the primordial atmosphere less efficiently \citep{Kegerreis2020,Denman2022}.

Giant impacts that induce strong turbulence also lead to atmospheric pollution by vaporizing rocky material.
In particular, head-on collisions with high impact velocities drive violent mixing between the primordial atmosphere and core material \citep[e.g.,][]{Kurosaki2023}.
The consumption of hydrogen by FeO in the rocky vapor forms a H$_2$--H$_2$O atmosphere at a high temperature of $\gtrsim$3000\,K.
Moreover, if the melt is not metal-saturated after a giant impact, the disk-oriented atmosphere can produce H$_2$O by reducing a molten mantle during the magma ocean period \citep{Lange1982,MatsuiAbe1986,Ikoma2006,Lupu2014,Itcovitz2022,Lichtenberg2022,Maurice2023}. 

The atmospheric mass and composition of a planet change during the planet's long-term evolution.
After a giant impact, an Earth-sized planet with H$_2$--H$_2$O atmosphere would be expected to be exposed to strong X-ray and extreme-UV (XUV) radiation from its host star.
An incident XUV flux drives the atmospheric escape from an Earth-sized planet \citep{Zahnle1988,Hamano2013}.
The hydrodynamic escape in the multi-component atmosphere is driven by major species.
As a result, the processes of XUV-driven mass loss lead to mass-dependent fractionation in the atmosphere \citep[e.g.,][]{1987Icar...69..532H,Johnstone2020,Yoshida2021,Yoshida2022}. 
If the hydrodynamic blow-off of light gases does not carry along heavy gases, the hydrodynamic escape of a planet with a H$_2$--H$_2$O atmosphere results in an increase of H$_2$O in the atmosphere.
Note that photochemical reactions via UV photons promote disequilibrium chemistry, such as the production of hydrocarbons, if carbon-bearing molecules such as CH$_4$ are abundant in the impact-generated atmosphere of an Earth-sized planet.

The phenomenon of energy-limited hydrodynamic escape has been well examined for sub-/super-Earths having H$_2$/He atmospheres \citep[e.g.,][]{Owen2013,Lopez2014,2020ApJ...889...77H} and H$_2$O-rich atmospheres \citep[e.g.,][]{Kurosaki2014}. 
Recently, \citet{Yoshida2022} calculated the mass loss of the H$_2$-H$_2$O atmosphere of an Earth-mass planet near the habitable zone around TRAPPIST-1-like stars coupled with chemical reactions and radiative cooling by H$_2$O, OH, H$^{+}_3$, and OH$^+$.
They found that the atmospheric escape rate of H$_2$ decreases as the initial H$_2$O/H$_2$ ratio increases.
However, it is not obvious that an Earth-mass planet can retain a water-rich atmosphere over Gyrs after giant impacts.
In this study, we calculate the long-term evolution of the water mass fraction in an impact-generated atmosphere of an Earth-mass planet around a Sun-like star under various initial conditions, such as different initial atmospheric masses and semimajor axes.
We consider the multi-component hydrodynamic escape of a H$_2$/He--H$_2$O atmosphere by photoevaporation under stellar XUV irradiation.
We also perform the Smoothed Particle Hydrodynamic (SPH) simulation of giant impacts on an Earth-mass planet with a primordial H$_2$/He atmosphere and then calculate chemical reactions between the remaining hydrogen and rocky vapor to determine the atmospheric compositions of the planet after giant impacts.

This paper is structured as follows. 
Section \ref{BS_U} describes our numerical models of mass loss from planetary atmospheres.
In Section \ref{sec:result}, we show the atmospheric compositions of an Earth-mass planet after giant impacts and the long-term atmospheric evolution through photoevaporation.
In Section \ref{sec:diss}, we calculate the chemical reaction between the rock vapor and the primordial atmosphere and show the SPH simulation results.
We also derive analytically the conditions necessary for the water-dominated atmosphere of an Earth-mass planet after the hydrodynamic escape.
We summarize our results in the last section.

\section{Model description} \label{BS_U}

\begin{figure}[htbp]
\begin{center}
\includegraphics[width=8cm]{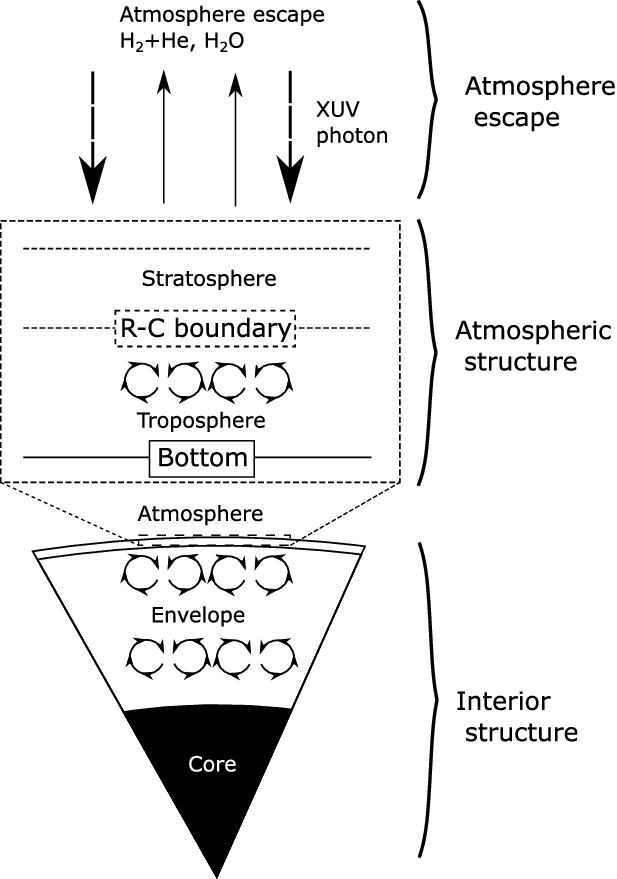}
\caption{Schematic of our planetary model. The planetary interior is composed of two parts: a spherical symmetry convective interior and a plane-parallel radiative-convective (R-C) equilibrium atmosphere. The interior structure in spherical symmetry is composed of a rocky core and H$_2$-He-H$_2$O envelope. The plane-parallel atmosphere is composed of H$_2$-He-H$_2$O (see also Section \ref{water}). \label{fig_interior}}
\end{center}
\end{figure}

We examine a one-dimensional, long-term evolution of an Earth-mass planet with the impact-generated atmosphere after giant impacts using the following numerical models. The planetary atmosphere is in radiative-convective equilibrium. \citep[see][for details]{Kurosaki2017}.
\begin{itemize}
\item The planetary interior is a one-dimensional, fully convective, layered core-envelope structure in hydrostatic equilibrium.
\item The planetary atmosphere is in a one-dimensional plain-parallel, radiative-convective equilibrium.
\end{itemize}
The following subsections summarize our models.
Figure~\ref{fig_interior} shows a schematic of our interior model.

\subsection{Interior structure}

We consider an Earth-mass planet that initially has a H$_2$-He atmosphere atop a rocky core.
Each layer in the interior is homogeneously mixed and fully convective.
We determine the interior structure by solving the following equations,
\begin{eqnarray}
\frac{\partial P}{\partial M_r} &=& -\frac{GM_r}{4\pi r^4}, \label{IE1_U}\\
\frac{\partial r}{\partial M_r} &=& \frac{1}{4\pi r^2\rho},\label{IE2_U} \\
\frac{\partial T}{\partial M_r} &=& -\frac{GM_r}{4\pi r^4}\frac{T}{P}\nabla, \label{IE3_U} \\
\frac{\partial L_r}{\partial M_r} &=& -T\frac{\mathrm{d}S}{\mathrm{d}t} \label{therm-1}, \label{eq:Lr}
\end{eqnarray}
where $r$ is the planetocentric distance, 
$M_r$ is the mass contained in the sphere of radius $r$, 
$P$ is the pressure, $\rho$ is the density, $T$ is the temperature, 
$G$ is the gravitational constant, 
$S$ is the specific entropy, $L_r$ is the total energy flux passing through a sphere of radius $r$, and $t$ is time.
The symbol $\nabla$ is the adiabatic temperature gradient with respect to pressure, namely,
\begin{equation}
\nabla = \nabla_{\mathrm{ad}} = \left(\frac{\partial \ln T}{\partial \ln P}\right)_S.
\end{equation}
As for the equations of state (EoS), we use \citet{Saumon1995} for hydrogen and helium and the M-ANEOS \citep{Thompson1972,Melosh2007,Thompson2019} of basalt for a rocky core \citep{Pierazzo2005}.
A giant impact onto an Earth-mass planet causes the dynamic mixing of a H$_2$-He atmosphere with vaporized core material.
Chemical reactions between hydrogen and rocky vapor produce hydrogen-bearing molecules, such as H$_2$O. We use the M-ANEOS of H$_2$O, which is one of the dominant components in the atmosphere after a giant impact (see Section \ref{water} for details).
We adopt the solar composition of the mass fraction of hydrogen $X=0.72$ and helium $Y=0.28$ as the initial envelope composition when the envelope has no rocky vapor \citep{Lodders2009}.
Integrating equation (\ref{eq:Lr}), we obtain the intrinsic luminosity $L_\mathrm{int}$ as
\begin{equation}
L_{\mathrm{int}}  =  - \left[ \frac{\mathrm{d}S_{\mathrm{env}}}{\mathrm{d}t} \int_{M_c}^{M_p} TdM_r  +\frac{\mathrm{d}S_{\mathrm{c}}}{\mathrm{d}t} \int_{0}^{M_c} TdM_r \right], \label{Eq_thermal_U}
\end{equation}
where $S_{\mathrm{env}}$ and $S_c$ are the specific entropies of the envelope and the core, respectively, 
$M_\mathrm{p}$ is the total mass of a planet,
and $M_c$ and $M_\mathrm{env}$ are the masses of the core and envelope, respectively. The intrinsic luminosity $L_{\mathrm{int}}$ is also written as $L_{\mathrm{int}}=4\pi R_p^2 F_{\mathrm{top}}$ where $R_p$ is the planetary radius and $F_{\mathrm{top}}$ is the outgoing flux from the top of the atmosphere \citep[e.g.,][]{Kurosaki2017}.

\subsection{Atmospheric structure}\label{sec:atom}

We assume a plane-parallel atmosphere in radiative-convective equilibrium.
As shown in Section \ref{water}, the rock-polluted atmosphere after a giant impact consists mainly of H$_2$, He, and H$_2$O.
The bottom of the atmosphere is assumed to be at $P_{\mathrm{btm}}$ = 1000\,bars.
Note that the condensation of molecular species occurs below $P_{\mathrm{btm}}$.
The temperature profile in the radiative atmosphere (i.e., the stratosphere) follows the analytical formula derived by \citet{MatsuiAbe1986}: 
\begin{equation}
\sigma T^4 =  F_{\mathrm{top}} \left( \frac{\tau + 1}{2} \right) + \frac{\sigma T_{\mathrm{eq}}^4}{2} \left[ 1+ \frac{\kappa_{\mathrm{th}}}{\kappa_{\mathrm{v}}} + \left( \frac{\kappa_{\mathrm{v}}}{\kappa_{\mathrm{th}}} - \frac{\kappa_{\mathrm{th}}}{\kappa_{\mathrm{v}}}  \right) e^{-\tau_{\mathrm{v}}} \right], \label{Eq_atm_str}
\end{equation}
where $F_\mathrm{top}$ is the net flux, $\sigma$ is the Stefan--Boltzmann constant,
$T_{\mathrm{eq}}$ is the equilibrium temperature, and $\kappa_{\mathrm{th}}$ (cm$^2\,$g$^{-1}$) and $\kappa_{\mathrm{v}}$ (cm$^2\,$g$^{-1}$) are the mean opacities for long- and short-wavelength radiation, respectively.
The optical depths for long- and short-wavelength radiation are denoted by $\tau$ and $\tau_{\mathrm{v}}$, respectively.
The optical depth $\tau$ is defined as $d\tau = -\rho\kappa_{\mathrm{th}} dr$.
Here we adopt the Rosseland opacity for $\kappa_{\mathrm{th}}$.
The H$_2$ and He opacities are calculated from the data table given in \citet{Freedman2008}.
The opacity of H$_2$O is calculated from line profiles \citep{Rothman1998} using the HITRAN 2020 database \citep{2022JQSRT.27707949G}.
We assume that $\kappa_{\mathrm{v}}=0.1\kappa_{\mathrm{th}}$ \citep{Guillot2010}. 
We consider the absorption of the radiation under the assumptions of zero reflectivity and scattering.
Then, $\kappa_\mathrm{th}$ in the H$_2$-He-H$_2$O atmosphere is given by
\begin{equation}
\kappa_\mathrm{th} = A_\mathrm{H_2} \kappa_{\rm H_2} + A_\mathrm{He}\kappa_{\rm He}  
+ A_\mathrm{W}\kappa_{\rm H_2O}, \label{kap_atm_nu}
\end{equation}
where $A_\mathrm{H_2}$, and $A_\mathrm{He}$, and $A_\mathrm{W}$ are the mole fractions of H$_2$, He, and H$_2$O, respectively.

The pseudo-moist adiabatic gradient determines the $T$-$P$ profile in the convective region (i.e., the troposphere). For $N$ kinds of species, including $j$ kinds of non-condensable ones, the pseudo-moist adiabatic temperature gradient \citep{Ingersoll1969,Atreya1986,Abe1988} is given by 
\begin{equation}
\left( \frac{d\ln T}{d\ln P} \right) = \nabla_\mathrm{dry}
\left[ \frac{1+\sum_{i=j+1}^{N} \frac{x_i}{1-x_{i}} \frac{d\ln~p_{i}^{\ast}}{d\ln~T} }
{1+\sum_{i=j+1}^{N} \frac{R_g}{C_p} \frac{x_i}{1-x_{i}} (\frac{d\ln~p_{i}^{\ast}}{d\ln~T})^2} \right], \label{Conv_atm_multi}
\end{equation}
where $\nabla_\mathrm{dry}$ is the adiabatic temperature gradient without condensation (i.e., the dry adiabatic), 
$C_p = \sum_{i=1}^{N} x_i C_{p,i} $ is the mean heat capacity, $A_i$ and $p_{i}^{\ast}$ are the mole fraction and vapor pressure of the $i$-th condensable species $(i = j+1,\cdots, N)$. 
The heat capacities of H$_2$O at constant pressure are $3.5R_g$
under the ideal gas approximation, where $R_g$ is the gas constant.
The non-ideal effect of the condensates is negligible near the photosphere.

We integrate the radiation transfer equations by using the Eddington approximation \citep[e.g.,][]{Abe1988}. The upward and downward radiation flux densities $F_{\mathrm{IR}}^{+}$ and $F_{\mathrm{IR}}^{-}$ can be written as
\begin{eqnarray}
F_{\mathrm{IR}}^{+}(\tau) &=& \pi B(\tau) \nonumber \\
&& - \int_{\tau_b}^{\tau} \frac{d}{d\tau'} ( \pi B(\tau') )
\exp\left[ -\frac{3}{2}(\tau' - \tau) \right] d\tau', \label{Fu} \\
F_{\mathrm{IR}}^{-}(\tau) &=& \pi B(\tau) \nonumber \\
&&- \int_{0}^{\tau} \frac{d}{d\tau'} ( \pi B(\tau') )
\exp\left[ -\frac{3}{2}(\tau-\tau') \right] d\tau' \nonumber \\
&&- \pi B(0) \exp\left( -\frac{3}{2}\tau \right), \label{Fd} 
\end{eqnarray}
and the net flux as
\begin{equation}
F_{\mathrm{net}} = F_{\mathrm{rad}}  + F_{c}  \label{Fnet},
\end{equation}
where $B (\tau)$ is the black body radiation intensity, $F_c$ is the convective flux\footnote{The convective flux is given by $F_\mathrm{C} = F_\mathrm{net}-F_\mathrm{rad}$.}, $F_{\mathrm{irr}}$ is the direct solar flux,
and $F_{\mathrm{rad}}$ is the net radiative flux given by $F_{\mathrm{rad}} = F_{\mathrm{IR}}^{+} - F_{\mathrm{IR}}^{-} - F_{\mathrm{irr}}$.
Note that $F_{\mathrm{top}}=F_{\mathrm{IR}}^{+} (\tau=0)$.
We assume that the net flux is constant at all altitudes.

\subsection{Atmospheric escape}\label{loss}
A planet undergoes atmospheric escape driven by a stellar X-ray and extreme UV (XUV) irradiation. The atmospheric mass loss rate is defined as
\begin{equation}
\left| \frac{dM_\mathrm{esc}}{dt} \right|  = \sum_{i=1}^{N} 4\pi R_p^2 m_i f_i, \label{FiMesc}
\end{equation}
where $M_\mathrm{esc}$ is the mass loss and $m_i$ and $f_i$ are the molecular weight and the escape flux for the $i-$th component, respectively.
The atmosphere of a planet after a giant impact is composed mainly of H$_2$, He, and H$_2$O.
Water molecules are transported upward in the atmosphere through diffusion processes. 
Hydrogen and helium can escape simultaneously because of the small mass difference between H and He, compared to O.
In this study, we consider hydrodynamic escape from a H$_2$-H$_2$O atmosphere, that is, under the assumption of $A_\mathrm{H} = A_\mathrm{H_2} + A_\mathrm{He}$.
Thus, we obtain the following approximate total escape flux:
\begin{equation}
\left| \frac{dM_\mathrm{esc}}{dt} \right| = 4\pi R_p^2 \left( m_{\mathrm{H}_2}^\ast f_\mathrm{H} + m_\mathrm{W} f_\mathrm{W} \right),
\label{eq:Mesc}
\end{equation}
where $m_\mathrm{H2}^\ast = m_\mathrm{H_2}A_\mathrm{H_2} + m_\mathrm{He} A_\mathrm{He}/(A_\mathrm{H_2} + A_\mathrm{He})$, the subscript $\mathrm{W}$ means water and $m$ is the molecular mass. 

The escape of H$_2$O is driven by collisions between hydrogen and water molecules.
Once water molecules are dredged up to the location where the photodissociation of H$_2$O occurs,
oxygen can escape if the escape flux of hydrogen is high enough to drag oxygen upward via collisions.
The upward transport of water molecules in the atmosphere controls the H$_2$O escape flux from the atmosphere.
When the flow of escaping hydrogen drags He and O up to a high altitude via collisions,
the H$_2$O escape flux \citep[e.g.,][]{Chamberlain1987} is described as
\begin{eqnarray}
f_\mathrm{W} = \left\{
\begin{array}{lc}
\frac{A_\mathrm{W}}{A_\mathrm{H}} f_\mathrm{H} - \frac{bg}{k_B T}A_\mathrm{W} \Delta m & (f_\mathrm{H} > f^\mathrm{crit}_\mathrm{H}), \\ \label{FW-def}  
    0 & (f_\mathrm{H} \le f^\mathrm{crit}_\mathrm{H}),
\end{array}
\right.
\end{eqnarray}
where
\begin{equation}
    f^\mathrm{crit}_\mathrm{H} = \frac{bg}{k_\mathrm{B}T}A_\mathrm{H}\Delta m, \label{fw_limit}
\end{equation}
$\Delta m = m_\mathrm{W} - m_{\mathrm{H}_2}$, $g$ is the gravitational acceleration, $T$ is the temperature, and $k_B$ is the Boltzmann constant.
The binary diffusion coefficient $b$ is written as
\begin{equation}
b = \frac{3}{64 Q} \sqrt{\frac{2\pi k_\mathrm{B} T (m^\ast_{\mathrm{H}_2}+m_\mathrm{W})}{ m^\ast_{\mathrm{H}_2} m_\mathrm{W} }}
\end{equation}
where $Q = \pi (\sigma_\mathrm{H} + \sigma_\mathrm{W})^2/16$ and $\sigma$ is the effective collision diameters of the molecules.
Combined with equations (\ref{eq:Mesc}) and (\ref{FW-def}), we find the H$_2$ escape flux as follows:
\begin{equation}
f_\mathrm{H} = A_\mathrm{H} \frac{\frac{1}{4\pi R_p^2}\left| \frac{dM_\mathrm{esc}}{dt}\right|+\frac{b g}{k_B T}m_\mathrm{W}A_\mathrm{W}\Delta m}{m^\ast_{\mathrm{H}_2}A_\mathrm{H} + m_\mathrm{W}A_\mathrm{W}}. \label{FH-def}
\end{equation}
We also obtain an upper limit to $A_\mathrm{H}$ for $f_\mathrm{W} >0$ using equation (\ref{eq:Mesc});
\begin{equation}
    A_\mathrm{H} < \frac{k_\mathrm{B}T}{4\pi G b m^\ast_{\mathrm{H}_2} \Delta m} \left|\frac{\dot{M}_\mathrm{esc}}{M_\mathrm{p}}\right|,
    \label{fw_limit2}
\end{equation}
where $M_\mathrm{atm}$ is the atmospheric mass and $\dot{M}_\mathrm{atm}/M_\mathrm{p} = dX_{\mathrm{H}}/dt$, assuming that $X_{\mathrm{H}}$ is the sum of the mass fraction of H$_2$ and He in the atmosphere.

When the atmospheric escape occurs in an energy-limited fashion, the total mass loss rate follows
\begin{equation}
\frac{dM_\mathrm{esc}}{dt} =-\frac{\varepsilon F_{\mathrm{XUV}}R_{p}\pi R_{\mathrm{XUV}}^2 }{GM_p K_{\mathrm{tide}}}, \label{ML01}
\end{equation}
where 
$\varepsilon$ is the XUV-heating efficiency in the upper atmosphere,
$F_{\mathrm{XUV}}$ is the incident XUV flux from the host star, 
$K_{\mathrm{tide}}$ is the potential energy reduction factor due to the stellar tide
and $R_{\rm{XUV}}$ is the effective radius at which the planet receives the incident XUV flux.
We use the formula for $K_{\mathrm{tide}}$ derived by \citet{Erkaev2007}
\begin{equation}
K_{\mathrm{tide}} = \frac{(\eta-1)^2(2\eta+1)}{2\eta^3}, \label{ML05}
\end{equation} where $\eta$ is the ratio of the Roche-lobe (or Hill) radius to the planetary radius, $R_p$.
In equation\,(\ref{ML01}), we assume $R_{\rm{XUV}}=R_{p}$, which is a good approximation for close-in planets of interest \citep{Lammer2013}.
It is noted that \citet{Lammer2013} focused on a H$_2$-He atmosphere. If a planet has a H$_2$O-rich atmosphere,
the scale height of the water vapor atmosphere is less than that of a H$_2$-He atmosphere at a given temperature. Nevertheless, $R_{\rm{XUV}}\simeq R_{p}$ is a good approximation even for the vapor atmosphere.

The XUV-heating efficiency in a non-H$_2$-He atmosphere remains poorly understood because of non-negligible contributions of minor gases such as CO$_2$ to radiative cooling. 
For the photoevaporation of hot Jupiters, $\varepsilon$ was estimated to be on the order of 0.1 (\citet{Yelle2008} and references therein). Thus, we adopt $\varepsilon=0.1$ as a fiducial value and investigate the sensitivity of our results to $\varepsilon$.
We suppose that the host star is a G-type star.
We adopt the empirical formula derived by \citet{Ribas2005} for $F_{\mathrm{XUV}}$ $\mathrm{erg}~\mathrm{s}^{-1}~\mathrm{cm}^{-2}$:
\begin{eqnarray}
F_{\rm{XUV}} = \left\{
\begin{array}{lc}
504 \left( \displaystyle \frac{a}{1\mathrm{AU}} \right)^{-2} &(t<0.1\rm{Gyr}) \\
29.7\left(\displaystyle \frac{t}{1\mathrm{Gyr}} \right)^{-1.23} \left( \displaystyle \frac{a}{1\mathrm{AU}} \right)^{-2} & (t\ge0.1\rm{Gyr}). \label{ML02}
\end{array}
\right.
\end{eqnarray}

\section{Results}\label{sec:result}
In this section, we present the long-term evolution of an Earth-mass planet with the impact-generated atmosphere after the giant impact.
First, we demonstrate SPH simulations of giant impacts that allow us to determine the mass fraction of rocky material in the remaining atmosphere in Section \ref{init}.
Second, we calculate chemical reactions between H$_2$-He gas in the remaining atmosphere and vaporizing rocky material. We show that an Earth-mass planet with an oxidizing core can have the H$_2$-He-H$_2$O atmosphere after the giant impact in Section \ref{water}. Third, we show the mass loss evolution of the H$_2$-H$_2$O atmosphere in Section \ref{sec:evolution}. Finally, we summarize results of long-term evolution of an Earth-mass planet with $10^{-3}$--10\,wt\% of the H$_2$-He-H$_2$O atmosphere at 0.1--3\,au after the giant impact in Section \ref{sec:gi}.

\subsection{Rock mass content in the atmosphere after a giant impact} \label{init}

In the atmosphere of an Earth-mass planet after a giant impact, the mixing ratio of the primordial H$_2$-He atmosphere to vaporized rocky material determines the initial H$_2$O content (see also Section\,\ref{water}).
To derive the water abundance in the atmosphere after a giant impact, we carried out SPH simulations of a head-on (non-oblique) collision 
with an Earth-mass planet with a 10\,wt\% H$_2$/He atmosphere.
The conditions for the impact simulations are as follows:
\begin{itemize}
    \item The atmospheric mass fraction is 1.0\% or 10\% by mass.
    \item The impact velocity is 1.0--2.6 times the mutual escape velocity. 
    \item A head-on collision is considered. 
    \item The impactor mass is 0.1 $M_\oplus$ or 1 $M_\oplus$.
\end{itemize}
In our SPH simulations, we consider that the inner part of the largest remnant whose mass fraction of rocky material is larger than 0.99 is the core-dominated region.
We estimate the atmospheric composition after a giant impact from the mass fractions of hydrogen and rocky material in the outer part above the ``core".

\begin{figure*}
\begin{center}
\includegraphics[width=\linewidth]{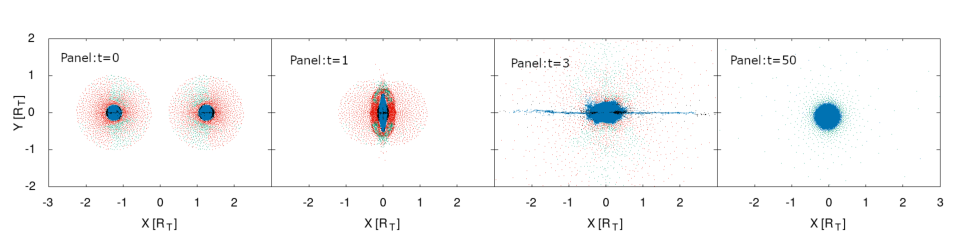}
\caption{Snapshots of the SPH simulation of a head-on collision with the impact velocity of the mutual escape velocity.
From left to right, snapshots show the time at $t=0, 1\tau_\mathrm{ff}, 3\tau_\mathrm{ff}$, and $50\tau_\mathrm{ff}$.
Both the target and the impactor have a $1M_\oplus$ core with 10\,wt\% of the H$_2$ atmosphere.
Green and blue particles denote the atmospheric and core components bound on the post-impact planet, respectively.
Red and black particles denote the escaping atmospheric and core components, respectively.
Both horizontal and vertical axes are normalized by the radius of the target, $R_{\rm T}$.
\label{sph_snap}}
\end{center}
\end{figure*}

Figure~\ref{sph_snap} shows an example of our giant impact simulations for $1~M_\oplus$ equivalent planets with a 10\,wt\% H$_2$/He atmosphere.
The impact velocity is equal to the mutual escape velocity.
First, the atmosphere flows out from the impact point just after the collision.
Once the core of the impactor collides with the core of the target, the atmosphere starts to eject from the antipode.
The ejection of the atmosphere occurs at the impact point during the oscillation of the merged core.
We check the mass fraction of rock to hydrogen after the impact.
The mixing profile is stable from 10 to 50 $\tau_\mathrm{ff}$, where $\tau_\mathrm{ff}$ is the free-fall timescale of the merged object. The loss fractions for the atmosphere and core are 84\% and 2.1\%, respectively.
The rock mass fraction in the atmosphere is 30\% by mass.
Our result suggests that the mixing state of the dredged-up rock vapor and atmospheric components is stable over the simulation timescale of this study because the rock vapor and atmosphere remain in hydrostatic equilibrium and are dynamically stable.

\begin{figure}
\begin{center}
\includegraphics[width=8cm]{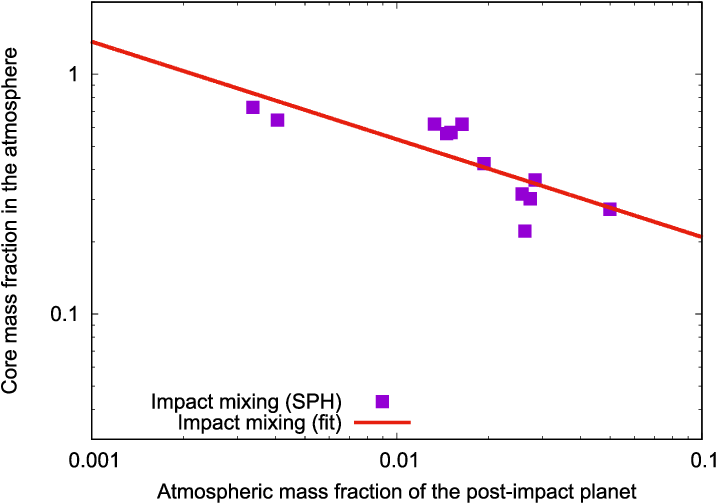}
\caption{
Mass fraction of core material in the remaining atmosphere ($Z_\mathrm{atm}$) of an Earth-mass planet after a giant impact. We consider head-on collisions between two Earth-mass planets.
The semi-empirical formula of the $Z_\mathrm{atm} - X_\mathrm{atm}$ relationship obtained from our SPH simulations of giant impacts,
$Z_\mathrm{atm}=8.2\times10^{-2}X_\mathrm{atm}^{-0.41}$
is also drawn (red line).
\label{MHa-mix}}
\end{center}
\end{figure}

We calculate the water mole fraction in the atmosphere of an Earth-mass planet after the giant impact under various impact conditions, using the mixing ratio of rocky vapor to hydrogen in the remaining atmosphere.
Figure\,\ref{MHa-mix}
shows the mass faction of core material ($Z_\mathrm{atm}$) in the atmosphere after a giant impact.
A more energetic giant impact can blow off more primordial H$_2$-He atmosphere and cause the effective mixing of H$_2$-He gas with rocky vapor.
Consequently, $Z_\mathrm{atm}$ increases with decreasing atmospheric mass fraction ($X_\mathrm{atm}$) after the impact.
Bearing in mind that the atmospheric loss induced by a giant impact should be controlled by the impact energy and energy loss due to a giant impact,
we obtain the following semi-empirical formula from our SPH simulations: 
$Z_\mathrm{atm}=8.2\times10^{-2}X_\mathrm{atm}^{-0.41}$.

\subsection{Atmospheric compositions after giant impacts} \label{water}

During a giant impact, vaporized rocky material reacts with the primordial atmosphere.
In this section, we focus on chemical reactions between a hydrogen-rich atmosphere and the core material vaporized by the shock of a giant impact.
We used \textsf{ggchem} \citep{Woitke2018} to calculate chemical reactions between the primordial H$_2$-He gas and rocky vapor, including dust condensation.
The primordial atmosphere has the same compositions as the protosolar abundances \citep{Asplund2021}.
We consider three kinds of core compositions: basalt \citep{Pierazzo2005}, CI chondrite \citep{Lodders2009}, and EH chondrite \citep{Javoy2010E&PSL.293..259J}.
The elemental abundances of basalt, CI chondrite, and EH chondrite are listed in Table~\ref{comp}.

\begin{table}[htbp]
    \centering
    \begin{tabular}{c|c|c|c|c}
     Element & Protosolar & Basalt & CI & EH  \\
     \hline
H   &  7.11E-01  &  -         &  1.96E-02  &  -         \\
He  &  2.71E-01  &  -         &  9.16E-09  &  -         \\
Li  &  9.89E-09  &  -         &  1.46E-06  &  -         \\
Be  &  1.76E-10  &  -         &  2.09E-08  &  -         \\
B   &  4.47E-09  &  -         &  7.74E-07  &  -         \\
C   &  2.86E-03  &  -         &  3.47E-02  &  -         \\
N   &  7.74E-04  &  -         &  2.94E-03  &  -         \\
O   &  6.41E-03  &  5.98E-01  &  4.58E-01  &  3.11E-01  \\
F   &  2.05E-07  &  -         &  5.81E-05  &  -         \\
Ne  &  1.90E-03  &  -         &  1.79E-10  &  -         \\
Na  &  3.11E-05  &  1.73E-02  &  4.98E-03  &  -         \\
Mg  &  7.09E-04  &  3.76E-02  &  9.67E-02  &  1.21E-01  \\
Al  &  5.93E-05  &  6.13E-02  &  8.49E-03  &  9.22E-03  \\
Si  &  7.50E-04  &  1.68E-01  &  1.06E-01  &  1.90E-01  \\
P   &  6.50E-06  &  -         &  9.66E-04  &  -         \\
S   &  3.47E-04  &  -         &  5.34E-02  &  -         \\
Cl  &  5.93E-06  &  -         &  6.97E-04  &  -         \\
Ar  &  7.18E-05  &  -         &  1.32E-09  &  -         \\
K   &  3.76E-06  &  -         &  5.43E-04  &  -         \\
Ca  &  1.73E-03  &  5.77E-02  &  9.21E-03  &  9.72E-03  \\
Sc  &  5.07E-08  &  -         &  5.89E-06  &  -         \\
Ti  &  3.67E-06  &  5.32E-03  &  4.50E-04  &  5.01E-04  \\
V   &  3.31E-07  &  -         &  5.42E-05  &  -         \\
Cr  &  1.79E-05  &  -         &  2.64E-03  &  3.61E-03  \\
Mn  &  1.18E-05  &  -         &  1.92E-03  &  -         \\
Fe  &  1.33E-03  &  5.27E-02  &  1.84E-01  &  3.31E-01  \\
Co  &  4.19E-06  &  -         &  5.05E-04  &  1.00E-03  \\
Ni  &  7.69E-05  &  -         &  1.07E-02  &  2.00E-02  \\
\hline
    \end{tabular}
    \caption{Elemental abundances of basalt, CI chondrite, and EH chondrite. The hyphen means the species that are not included in our composition models.}
    \label{comp}
\end{table}

\begin{figure}
\begin{center}
\includegraphics[width=8cm]{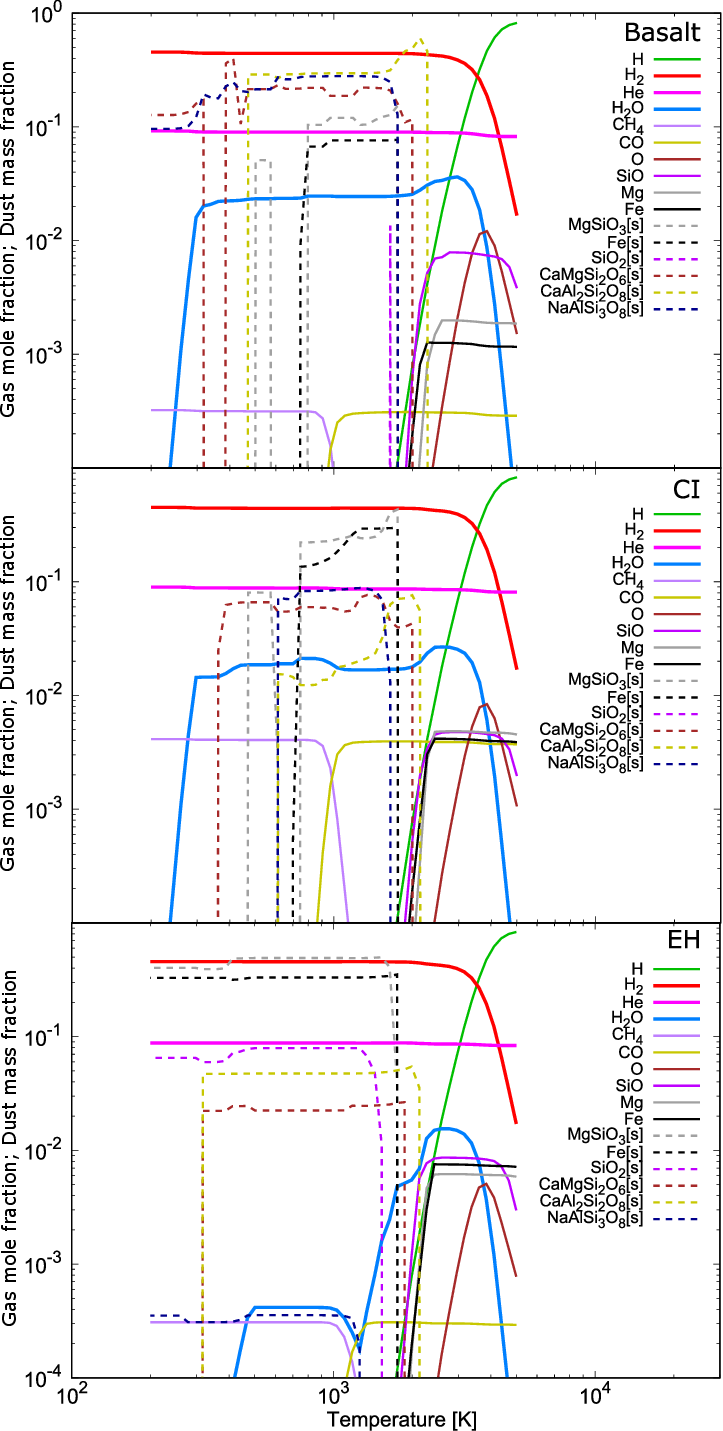}
\caption{Major species in an 1\,bar atmosphere of an Earth-mass planet with three types of core material after a giant impact: basalt (top panel), CI (middle panel), and EH (bottom panel).
We assumed that the mixing ratio of rocky vapor to the primordial atmosphere was 50\%.
Solid and dashed lines show the mole fractions of atomic and molecular species in the gas phase and the mass fractions of dust species, respectively.
\label{ggchem_T}}
\end{center}
\end{figure}

\begin{figure}
\begin{center}
\includegraphics[width=8cm]{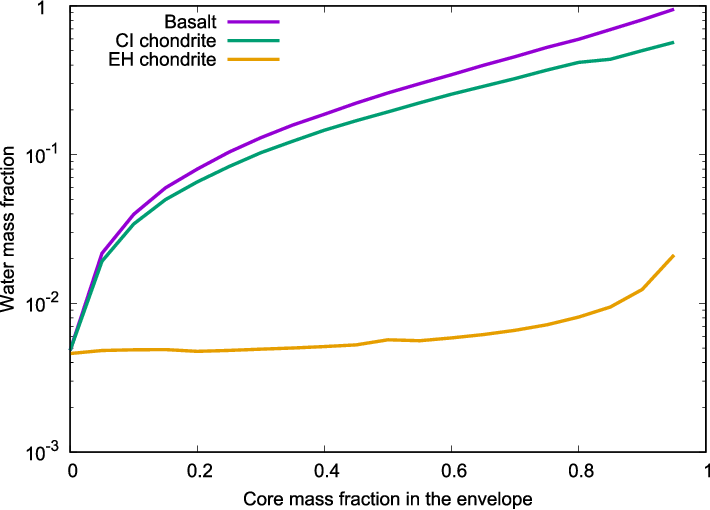}
\caption{
Mass fractions of H$_2$O at $T = 1000$~K in the rock-polluted atmosphere after a giant impact. 
The purple, green, and orange lines correspond to the case of basalt and CI chondrite, and EH chondrite as core material, respectively.
}
\label{H2H2O}
\end{center}
\end{figure}

Figure~\ref{ggchem_T} shows the mole fractions of major atomic and molecular species in the primordial atmosphere uniformly mixed with rocky vapor, where the mixing ratio is assumed to be 50\%.
We also show the mass fractions of six condensates such as MgSiO$_3$ (enstatite), CaMgSi$_2$O$_6$ (diopside), CaAl$_2$Si$_2$O$_8$ (anorthite), and NaAlSi$_3$O$_8$ (albite).
The atmospheric compositions of an Earth-mass planet in chemical equilibrium have two kinds of regimes: 
(i) a mineral atmosphere composed of vaporized metal atoms and metal oxides at a high temperature of $\gtrsim$3000\,K;
and (ii) a volatile-rich atmosphere at a lower temperature.
The major components in the low-$T$ atmosphere are H$_2$, He, and H$_2$O. 
Water molecules are produced by reducing FeO by hydrogen at $T \lesssim 3000$\,K.
The abundance of H$_2$O is higher than that of CO or CH$_4$ by two orders of magnitude when the core is composed of basalt or CI chondrite. For the EH-chondrite-like core, the H$_2$O abundance is comparable to that of CO or CH$_4$ because oxygen is distributed to dust particles.
An Earth-mass planet is likely to have a mineral atmosphere right after a giant impact.
As the planet rapidly cools, its atmosphere becomes H$_2$O-rich because all the metals become dust below $\sim$3000\,K and settle into the interior.
The mass fractions of H$_2$ and H$_2$O vary with the mixing ratio of rocky vapor ($Z_\mathrm{atm}$) in the atmosphere of an Earth-mass planet (see Figure~\ref{H2H2O}). 
As the $Z_\mathrm{atm}$ in the atmosphere becomes high,
H$_2$ in the atmosphere is preferentially replaced for H$_2$O. 
We also find that the mass fraction of H$_2$O strongly depends on the core compositions. The water fraction in the impact-generated atmosphere becomes high unless an Earth-mass planet has a reducing core such as EH chondrites.

\subsection{Mass loss of the H$_2$--H$_2$O atmosphere} \label{sec:evolution}

In this subsection, we discuss the long-term evolution of a H$_2$-H$_2$O atmosphere.
The upward transport of H$_2$O (i.e., $f_\mathrm{W} > 0$) in the atmosphere via collisions between hydrogen and water occurs while $A_\mathrm{H}$ satisfies equation (\ref{fw_limit}).
The planet can retain the initial H$_2$O content if the escape of H$_2$O is suppressed by inefficient diffusion processes in the atmosphere.

\begin{figure}
\begin{center}
\includegraphics[width=8cm]{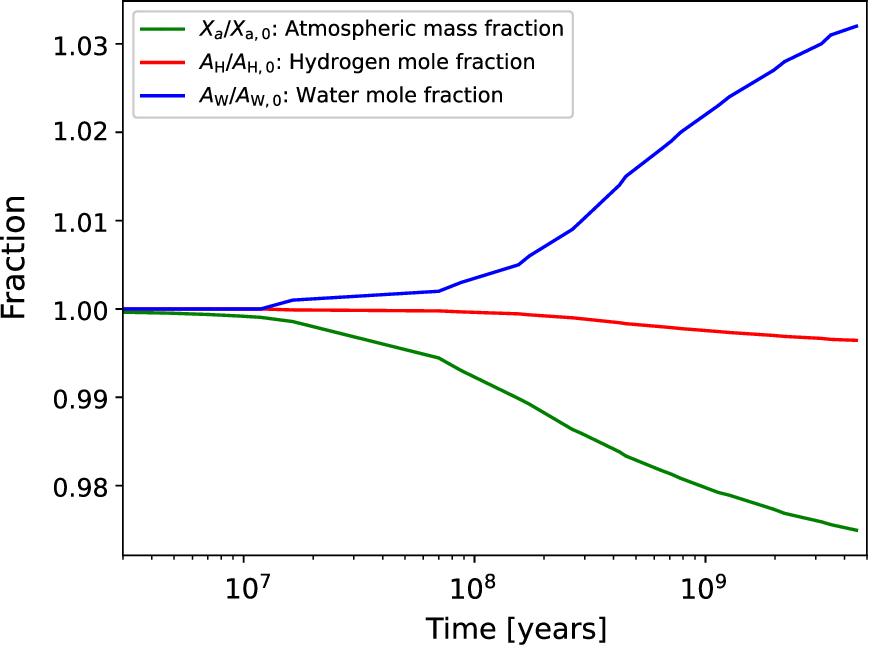}
\caption{
Mole fractions of hydrogen + helium (red line) and water (blue line) in the evaporating atmosphere after a giant impact. 
We considered an Earth-mass planet that initially has a 6.3\,\% H$_2$-H$_2$O atmosphere with an initial water mole fraction of $A_\mathrm{W} = 10\,\%$, which is equivalent to a mixing ratio of rocky vapor to H$_2$-He gas of 90\,mol\,\%, at $a=1$\,au around a Sun-like star.
The planet loses its atmosphere through photoevaporation ($\varepsilon = 0.1$) for 4.6\,Gyrs.
The green line plots the atmospheric mass fraction. 
The non-smooth features of the blue line is due to the sparse table of H$_2$O EoS.
\label{ME}}
\end{center}
\end{figure}

Figure\,\ref{ME} demonstrates the atmospheric evolution of an Earth-mass planet initially having a 6.3\,wt\%  H$_2$-H$_2$O atmosphere with a water mole fraction of $A_\mathrm{W} = 10\%$ at $a=1$\,au around a Sun-like star.
After a giant impact, the planet cools rapidly and reaches the radiative equilibrium state\footnote{The radiative cooling of a planet is characterized by the Kelvin-Helmholtz timescale;
\begin{equation}
    \tau_\mathrm{KH} = \frac{GM_\mathrm{p}^2}{2R_\mathrm{p}L} \sim 10^{3} \left( \frac{M_\mathrm{p}}{1\,M_\oplus} \right) \left( \frac{R_\mathrm{p}}{2\,R_\oplus} \right)^3 \left( \frac{T}{3000\,\rm K} \right)^{-4}~\mathrm{yrs},
\end{equation}
where $L$ is the luminosity. A hot, inflated Earth-mass planet gravitationally contracts for $\sim$10$^3$\,yrs.
}.
The atmospheric escape occurs in two stages in $\gtrsim$10\,Myrs.
Both hydrogen and water molecules escape as long as the escape flux of hydrogen is high enough to drag oxygen.
After $\sim$1000\,Myrs, less heavy hydrogen preferentially escapes from the atmosphere by the XUV-driven photoevaporation, whereas H$_2$O molecules stay in the lower atmosphere. Consequently, as the atmospheric mass decreases, the mole fraction of heavier molecules (H$_2$O) apparently increases.
Thus, the final state of the planetary atmosphere should be enriched in water molecules. 
In this simulation, the final atmospheric mass fraction is 6.15\%, which includes $A_{\rm H} = 89.7$ mol\% and $A_{\rm W} = 10.3$ mol\%.

\subsection{Long-term evolution of the impact-generated atmosphere formed by a giant impact} \label{sec:gi}

In Section~\ref{init}, we obtained a semi-empirical scaling law for the initial rock vapor mass fraction in the impact-generated atmosphere (see Figure\,\ref{MHa-mix}).
The water mass fraction ($X_\mathrm{W}$) is determined as a function of $X_\mathrm{atm}$ shown in Figure~\ref{H2H2O}.
We perform the same long-term simulations of an Earth-mass planet with the H$_2$-H$_2$O atmosphere using the scaling law as those in Section \ref{sec:evolution}.
We discuss how the results of Figure~\ref{AH1} change if the $X_\mathrm{W}$ is used as the initial condition of the impact-generated atmosphere of an Earth-mass planet. Note that $X_\mathrm{W}$ is given by $Z_\mathrm{atm}$ shown in figure~\ref{H2H2O} which is given by the scaling relation of  $Z_\mathrm{atm} = 8.2 \times 10^{-2} X_\mathrm{atm}^{-0.41}$.

\begin{figure*}
\begin{center}
\includegraphics[width=\linewidth]{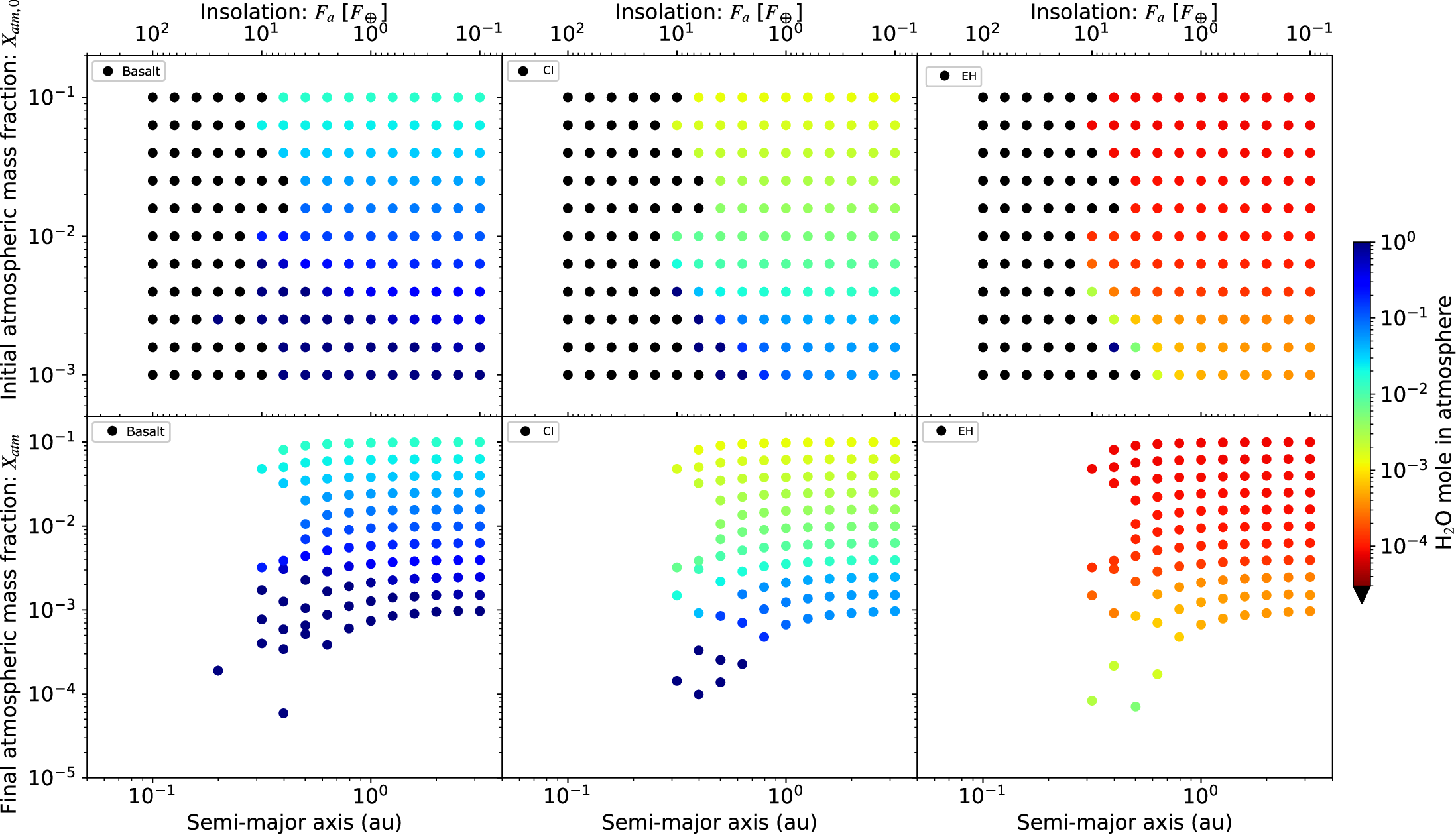}
\caption{Initial and final atmospheric mass fraction $X_\mathrm{atm}$ and final water mole fraction (color bar) after the atmospheric loss through photoevaporation ($\varepsilon = 0.1$) for 4.6\,Gyrs. 
The rocky core is composed of basalt (left panel), CI (middle panel), and EH (right panel).
An initial water mass fraction is determined by the empirical relation derived from the giant impact simulations, $X_\mathrm{W} = 8.2 \times 10^{-2} X_\mathrm{atm}^{-0.41}$.
The color contour shows the final H$_2$O mole fraction in the remaining atmosphere.
\label{AWSPH}}
\end{center}
\end{figure*}

The left, middle, and right panels of Figure~\ref{AWSPH} show the relationship between
the initial ($X_{\mathrm{atm},0}$: top) and final ($X_{\mathrm{atm}}$: bottom) atmospheric mass fraction of an Earth-mass planet and
the water mole fraction in the remaining atmosphere after the atmospheric loss through photoevaporation over 4.6\,Gyr for core models with the basalt, CI, and EH compositions, respectively. 
A water-dominated atmosphere is more likely to form in a less massive atmosphere after a giant impact;
the mole fraction of water remaining in the atmosphere exceeds 10\,\% when $X_{\mathrm{atm}} \leq 4 \times 10^{-3}$ for the case of a basalt core model and $X_{\mathrm{atm}} \leq \times 10^{-3}$ for the case of a CI-chondritic core one.
In contrast, such a water-rich atmosphere is difficult for an Earth-mass planet with a reducing core like an EH chondrite to form.
We find that an Earth-mass planet requires an oxidizing core to produce a water-rich atmosphere in the giant impact scenario, followed by the atmospheric escape.

\section{Discussion}\label{sec:diss}

We see the dependence of the final atmospheric water fraction on the semi-major axis and the initial water content 
 (Section\,\ref{param_std}). We also present the analytical condition for the water-dominated atmosphere (Section \ref{cond-water}), and discuss the caveats of our models (Section \ref{sec:caveat}).

\subsection{Dependence of the final atmospheric water fraction on the semi-major axis and the initial water content} \label{param_std}

\begin{figure*}
\begin{center}
\includegraphics[width=\linewidth]{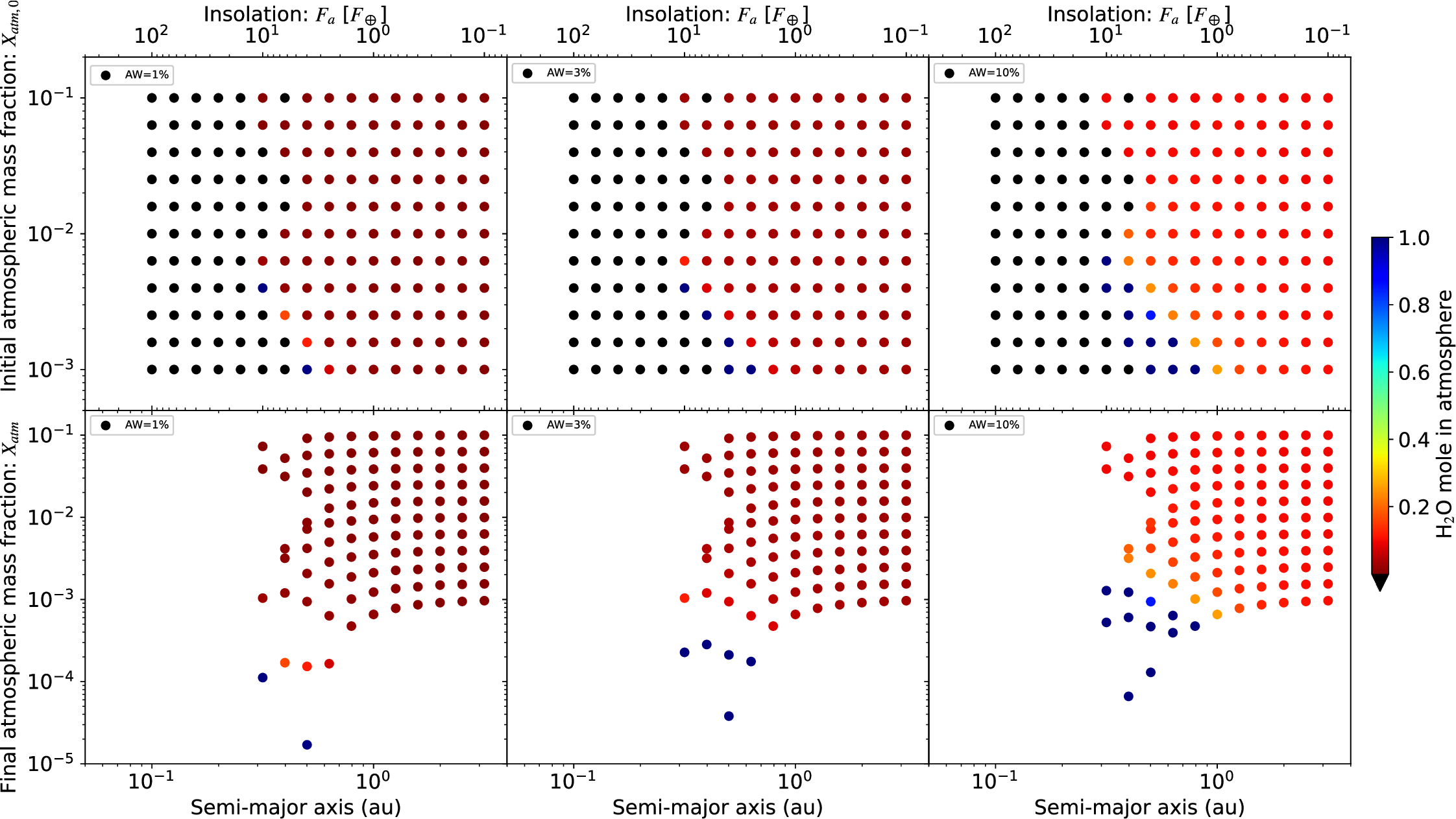}
\caption{
Same as Figure\,\ref{AWSPH} but with different initial water mole fractions (1\%, 3\%, and 10\% in the left, middle, and right panels, respectively). The rocky core is assumed to be basalt in all simulations.
\label{AH1}}
\end{center}
\end{figure*}

We examine how mass loss affects the water fractions in the atmospheres of Earth-mass planets after giant impacts with various semimajor axes and initial water content. We consider that the initial atmospheric mass fraction relative to the total mass of a planet ranges from $ 10^{-3}$ to $ 10^{-1}$. The initial H$_2$O mole fractions in the atmosphere $A_{\mathrm W}$ are assumed to be 1\%, 3\%, and 10\%.
Figure\,\ref{AH1} shows the initial and final atmospheric mass fractions and the final water mole fraction in the atmosphere of an Earth-mass planet with a given equilibrium temperature after $t=4.6~$\,Gyrs.

There are three key parameters for the survival of the water-rich atmosphere of an Earth-mass planet: the semimajor axis of the planet, $a$, the initial water mole fraction in the atmosphere, $A_\mathrm{W}$, and the initial atmospheric mass, $X_\mathrm{atm}$.
When the planet is close to its host star, it loses almost all of its H$_2$-H$_2$O atmosphere and becomes a bare rocky planet.
If the planet is distant from its star and exposed to less intensive XUV irradiation, neither its atmospheric mass $X_{\mathrm{atm}}$ nor its water mole fraction $A_\mathrm{W}$ changes significantly.
In this case, the final water mole fraction in the atmosphere reflects the initial water content that the planet has formed.
Water molecules tend to remain in the atmosphere if the atmospheric mass fraction is less than about 0.3\%.
The initial atmosphere with a higher initial $A_\mathrm{W}$ is likely to survive because a low escape flux of hydrogen in the water-rich atmosphere reduces the efficiency of the upward transport of water molecules.

The XUV-driven atmospheric escape of an Earth-mass planet yields a high H$_2$O abundance in the atmosphere for planets with $X^{\mathrm{initial}}_\mathrm{atm} \lesssim$ 0.3\% ($X^{\mathrm{final}}_\mathrm{atm} <$ 0.1\%)  when it receives $F_\mathrm{a} > 2F_\oplus$. 
Such a planet should have a steam atmosphere because its insolation exceeds the critical flux for a runaway greenhouse effect around a Sun-like star \citep[e.g.][]{2014ApJ...787L..29K,2019JGRE..124.2306K}.
The less efficient heating of hydrogen by UV photons (e.g., $\varepsilon = 0.01$) moves the outer edge of a ``steamy zone” closer to a central star (see also Figure \ref{watermass}).

One of the key findings of the present study is that the enhancement of H$_2$O in the atmosphere of an Earth-mass planet occurs if $f_\mathrm{W}=0$, i.e., in the case that only hydrogen escapes.
Unless a planet loses its atmosphere completely, a low escape flux of hydrogen in a less massive H$_2$--H$_2$O atmosphere that satisfies $f_\mathrm{W}=0$ forms a water-dominated atmosphere in Gyrs.
In Section~\ref{cond-water}, we analytically derive the condition for the formation of a water-dominated atmosphere of an Earth-mass planet.
Also, as we will see in Section~\ref{sec:gi}, even when the results of giant impact simulations are taken into account, there is no change in this finding.

\subsection{Analytical condition for the water-dominated atmosphere} \label{cond-water}
In Section \ref{sec:evolution}, we demonstrate that the Earth-mass planet with a small atmospheric mass fraction at $a = 1$\,au has a water-dominated atmosphere in Gyrs.
We analytically derive the condition in which Earth-mass planets have water-dominated atmospheres.

The outflow of escaping hydrogen cannot drive the escape of water molecules if $f_\mathrm{W} = 0$, that is, $f_\mathrm{H} \leq f^{\mathrm{crit}}_\mathrm{H}$
(see equation (\ref{fw_limit})).
The mass loss process increases the H$_2$O mole fraction when $f_\mathrm{H} \leq f^{\mathrm{crit}}_\mathrm{H}$.
Integrating equation (\ref{fw_limit2}) from $t=0$ to $t = \Delta t$, we obtain the upper-limit mass fraction of H$_2$ and He
in the impact-generated atmosphere in which all water molecules can persist for $\Delta t$;
\begin{equation}
    X^{\mathrm{max}}_{\mathrm{H}} \equiv \frac{M_\mathrm{esc}}{M_\mathrm{p} (t = 0)}
        = \frac{4\pi G b m_{\mathrm{H}_2} \Delta m}{k_\mathrm{B} T} A_\mathrm{H}(t=0) \Delta t,
\label{eq:Xa_crit}
\end{equation}
where $M_\mathrm{esc}$ is the net mass loss of a H$_2$-He atmosphere.
This value is given by $M_\mathrm{esc} = \int \frac{dM_\mathrm{esc}}{dt} dt \approx  3\varepsilon \bar{F}\Delta t/(4G \bar{\rho}_\mathrm{p})$, where $\bar{F}=\int F_\mathrm{XUV} dt$ and $\bar{\rho}_p$ is the planetary mean density.
An Earth-mass planet can have a water-dominated atmosphere if it initially has an impact-generated atmosphere of $ X_{\mathrm{H}} \leq X^\mathrm{max}_{\mathrm{H}}$.
The condition for a water-dominated atmosphere of a planet, $X^{\mathrm{max}}_{\mathrm{H}}$, well explains our numerical results as shown in Figure\,\ref{AH1}.
In fact, Earth-mass planets with water-dominated atmospheres appear just below the criterion of $X^{\mathrm{max}}_{\mathrm{H}}$ after 4.6\,Gyrs unless such planets lose their atmospheres completely by photoevaporation after a giant impact. 

Next, we consider the range of semimajor axes in which Earth-mass planets with water-dominated atmospheres can exist.
We can obtain the approximate semimajor axis within which a planet loses all of its the H$_2$/He, $a_\mathrm{esc}$: 
\begin{equation}
    a_\mathrm{esc} = \left(\frac{3}{4G}\frac{\overline{\epsilon L_\mathrm{XUV}} \Delta t}{M_\mathrm{p} \rho_p X_\mathrm{atm}} \right)^{\frac{1}{2}} \label{aesc-def},
\end{equation}
where $\Delta t$ is the time elapsed after a giant impact, $M_\mathrm{atm}$ is the atmospheric mass of a planet, $\rho_\mathrm{p}$ is the mean density of a planet, and $\overline{\epsilon L_\mathrm{XUV}}$ is the time-averaged XUV luminosity that a planet receives. Note that $K_\mathrm{tide} \sim 1$ for an Earth-mass planet in this study.
A large hydrogen flux causes the escaping flow of water molecules, namely when $f_\mathrm{H} > f^{\mathrm{crit}}_\mathrm{H}$ (see equation (\ref{fw_limit})).
The farther a planet is from a central star, the lower the escape rate of hydrogen becomes under stellar XUV irradiation. The critical semimajor axis ($a_\mathrm{W}$) beyond which an Earth-mass planet prevents the threat of H$_2$O escape is defined by $f_\mathrm{H} = f^{\mathrm{crit}}_\mathrm{H}$;
\begin{eqnarray}
    a_\mathrm{W} = \sqrt{\frac{3}{4G}\frac{\overline{\epsilon L_\mathrm{XUV}} }{\rho_p} \frac{k_B T_\mathrm{atm}}{4\pi m_H^2 A_H G M_p b(T_\mathrm{atm})}} \label{aw-def},
\end{eqnarray}
where $T_\mathrm{atm}$ is the atmospheric temperature.
If $a_\mathrm{W} < a < a_\mathrm{esc}$, the planet undergoes the hydrodynamic escape of hydrogen, but it preserves the initial amount of water in the atmosphere.

\begin{figure}
\begin{center}
\includegraphics[width=8cm]{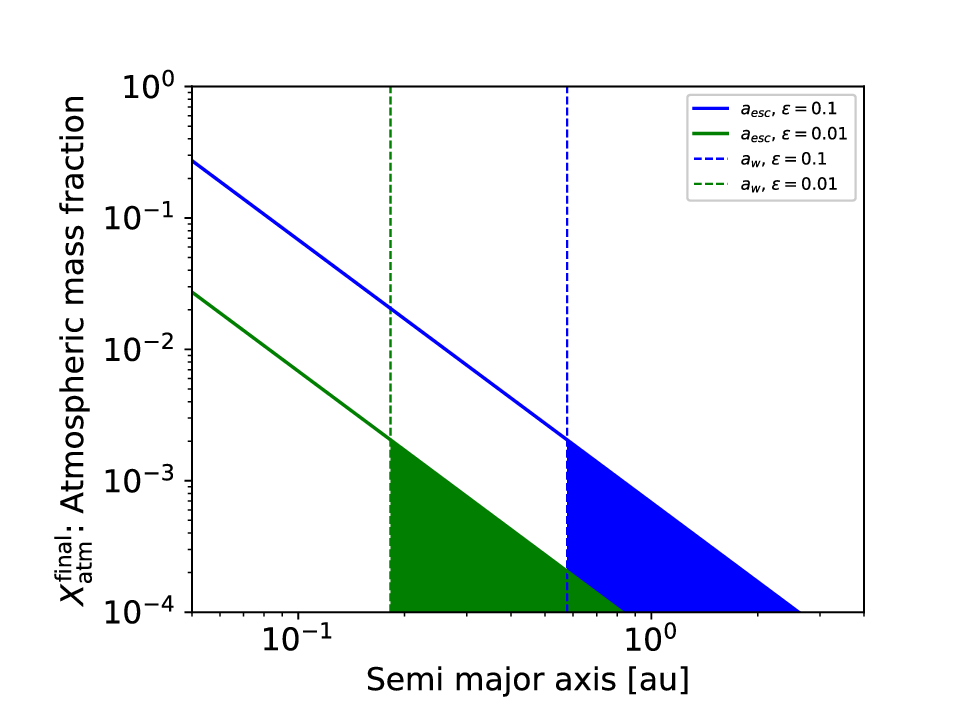}
\caption{
Habitat of an Earth-mass planet with a water-rich atmosphere that orbits a Sun-like star.
Dashed and solid lines represent $a_\mathrm{esc}$ (Eq. \ref{aesc-def}) and $a_\mathrm{W}$ (Eq. \ref{aw-def}) in two models with $\varepsilon = 0.1$ (blue) and $0.01$ (green), respectively.
Blue and green shaded regions show the conditions where an Earth-mass planet can retain water in the atmosphere for Gyrs.
\label{watermass}}
\end{center}
\end{figure}

Figure~\ref{watermass} shows the habitat of an Earth-mass planet that has a water-dominated atmosphere (i.e., $a_\mathrm{W} < a < a_\mathrm{esc}$, hereafter the steamy zone").
We set $\overline{L_\mathrm{XUV}} = 3.1\times 10^{28}~\mathrm{erg}\cdot\mathrm{s}^{-1}$ for a Sun-like star and $T_\mathrm{atm}=500~$K.
Since the atmospheric temperature at the upper stratosphere is higher than the equilibrium temperature, we assume that $T_\mathrm{atm}$ is determined by the planet's atmosphere at typically $10^8$ years old.
The XUV heating efficiency, $\varepsilon$, determines the inner and outer edges of a steamy Earth-mass planet.
The formation of a steam atmosphere requires the initial atmosphere mass fraction $X_\mathrm{atm} \lesssim 2\times10^{-3}$ regardless of $\varepsilon$.
These ``steamy zones" are in good agreement with the results shown in Figure \ref{AH1}.

\subsection{Caveats}\label{sec:caveat}

This study considered a head-on giant impact on an Earth-mass planet. In reality, grazing impacts occur in the final stage of terrestrial planet formation \citep[e.g.,][]{2010ApJ...714L..21K,2020ApJ...899...91O}. Less efficient mixing of a H$_2$/He atmosphere and vaporized core material by oblique collisions decreases the production of H$_2$O.
We also assumed simple rock compositions for core material.
If the core of an Earth-mass planet is a mixture of icy and rocky material, the chemical compositions of an impact-generated atmosphere become different.
Moreover, dust clouds may form in the polluted atmosphere unless dust particles fall into the deep interior.

The atmospheric compositions generated by the impact are still debated. \cite{Lupu2014} showed that a post-impact planetary atmosphere is dominated by CO$_2$ as well as H$_2$O.
We should underestimate the abundance of carbon-bearing species since the rock material is assumed to be carbon-depleted. The mass loss efficiency in our study may be overestimated because CO$_2$ helps to cool the upper atmosphere \citep{Yoshida2020}.
Another caveat involves the mixing rate of rock material in the primordial atmosphere.
We considered only head-on collisions, which cause the most efficient mixing between the atmosphere and the rock vapor, as indicated by the Moon-forming impact simulations on the proto-Earth \citep{Nakajima2015}.
The present study focused on the most optimistic case, where a giant impact into an Earth-sized planet forms a water-rich atmosphere.
Detailed atmospheric compositions of Earth-sized planets after a non-oblique giant impact will be discussed in our future work.

\section{conclusion}
We showed that the mixing of hydrogen in the atmosphere and rocky material vaporized by a giant impact into an Earth-sized planet results in the production of H$_2$O via chemical reactions.
We simulated the long-term evolution of the H$_2$-H$_2$O atmospheres of Earth-mass planets around a Sun-like star for Gyrs after giant impacts.
We examined the mass loss from an Earth-sized planet that initially has a H$_2$-H$_2$O atmosphere of 0.1\,wt\% -- 10\,wt\% at a semimajor axis from 0.1 to 3\,au.
Our key findings are summarized as follows.
\begin{itemize}
    \item The composition of the impact-generated atmosphere is dominated by H$_2$, He, and H$_2$O when it has cooled below 3000 K.
    \item Earth-mass planets can avoid water loss through photoevaporation if the escape flux of hydrogen is too low to drag oxygen via collisions.
    \item The post-impact atmosphere of an Earth-mass planet evolves into a water-dominated atmosphere if its atmospheric mass fraction is less than a few times 0.1\% provided that the core is oxidizing material such as the basalt or CI chondrite.
    \item If the rocky core is composed of a reducing material such as EH chondrite, the post-impact atmosphere should not be water-dominated, even if the impact-induced mixing between H$_2$-He gas and rocky vapor is efficient.
\end{itemize}

 An Earth-mass planet in a habitable zone can retain the initial amount of H$_2$O in its atmosphere if $X_\mathrm{atm}^{\mathrm{final}} \lesssim 10^{-3}$ after a giant impact, although part of the remaining water exists as ``oxygen" in the atmosphere due to photodissociation. 
 Note that the water-rich atmosphere of an Earth-mass planet can be found on the core composed of basalt and CI chondrite in the giant impact scenario, while the water-poor atmosphere forms on the core composed of EH chondrite.
The advanced capability of the James-Webb space telescope (JWST) allows it to observe the atmospheres of Earth-sized planets around nearby stars, e.g., TRAPPIST-1 planets \citep{2019AJ....158...27L}. In the late 2020s, {\it PLATO} \citep{2014ExA....38..249R}, \textit{Earth 2.0} \citep{2022arXiv220606693G}, and the Roman space telescope \citep{2019ApJS..241....3P} are expected to find hundreds of Earth-sized planets in a habitable zone around Sun-like stars and M-dwarfs. 
Terrestrial planets with water-dominated atmospheres, as considered in this study, should be interesting targets for JWST observations.
As shown in Figure~\ref{AWSPH}, such terrestrial planets can exist in orbits of semimajor axes ranging from a few 0.1 au to a few au, depending on the photoevaporation efficiency ($\varepsilon$).
This constraint on the locations of terrestrial planets with water-dominated atmospheres helps elucidate the mass-loss efficiency.

\begin{acknowledgments}
We thank the anonymous referee for improving our manuscript.
K.K. is supported by JSPS KAKENHI Grants-in-Aid for Scientific Research No. 20J01258, 21H00039, 23H01231.
Y.H. is supported in part by a JSPS KAKENHI Grant Numbers 18H05439.
M.O. is supported by JSPS KAKENHI Grant Numbers JP18K13608 and JP19H05087.
Numerical computations of SPH simulations were carried out on a Cray XC50 at the Center for Computational Astrophysics, National Astronomical Observatory of Japan.
\end{acknowledgments}

%

\vspace{5mm}


\software{GGchem \citep{Woitke2018}}








\bibliography{sample631}{}
\bibliographystyle{aasjournal}

\end{document}